\documentclass[aps,twocolumn, superscriptaddress, 10pt,floatfix,showpacs]{revtex4-1}

\usepackage{graphicx}
\usepackage{color}
\usepackage{amsfonts, amsmath, amsthm, amscd}

\usepackage{tensor}
\usepackage{braket}

\begin{document}

\title{Competing interactions in semiconductor quantum dots}

\author{R. van den Berg}
\email{R.vandenBerg2@uva.nl}
\affiliation{Institute for Theoretical Physics, University of Amsterdam, Science Park 904\\
Postbus 94485, 1090 GL Amsterdam, The Netherlands}

\author{G.P. Brandino}
\affiliation{Institute for Theoretical Physics, University of Amsterdam, Science Park 904\\
Postbus 94485, 1090 GL Amsterdam, The Netherlands}

\author{O. El Araby}
\affiliation{Institute for Theoretical Physics, University of Amsterdam, Science Park 904\\
Postbus 94485, 1090 GL Amsterdam, The Netherlands}

\author{R.M. Konik}
\affiliation{CMPMS Dept. Bldg 734 Brookhaven National Laboratory, Upton NY 11973, USA}

\author{V. Gritsev}
\affiliation{Institute for Theoretical Physics, University of Amsterdam, Science Park 904\\
Postbus 94485, 1090 GL Amsterdam, The Netherlands}

\author{J.-S. Caux}
\affiliation{Institute for Theoretical Physics, University of Amsterdam, Science Park 904\\
Postbus 94485, 1090 GL Amsterdam, The Netherlands}
\date{\today}

\begin{abstract}
We introduce an integrability-based method enabling the study of semiconductor quantum dot models incorporating both the full hyperfine interaction as well as a mean-field treatment of dipole-dipole interactions in the nuclear spin bath. By performing free induction decay and spin echo simulations we characterize the combined effect of both types of interactions on the decoherence of the electron spin, for external fields ranging from low to high values.
We show that for spin echo simulations the hyperfine interaction is the dominant source of decoherence at short times for low fields, and competes with the dipole-dipole interactions at longer times. On the contrary, at high fields the main source of decay is due to the dipole-dipole interactions. In the latter regime an asymmetry in the echo is observed. Furthermore, the non-decaying fraction previously observed for zero field free induction decay simulations in quantum dots with only hyperfine interactions,  is destroyed for longer times by the mean-field treatment of the dipolar interactions.
\end{abstract}

\maketitle

\section{Introduction}
One of the great challenges in quantum dot experiments is to gain control over decoherence effects due to the presence of nuclear spins in the underlying substrate. 
Motivated by these experimental difficulties, numerous theoretical studies have been dedicated to explaining the influence of different aspects of the nuclear environment of a quantum dot
\cite{2002_Khaetskii_PRL_88,2002_Schliemann_PRB_66,2003_Schliemann_JPhys,2003_Sousa_PRB_67,2003_Sousa_PRB_68,2004_Coish_PRB_70,2005_Shenvi_PRB_71,2005_Sousa_PRB_72,
2005_Witzel_PRB_72,2007_Saikin_PRB_75,2009_Cywinski_PRB_79,2009_Cywinski_PRL_102,2010_Cywinski_PRB_82,2009_Tribollet_EurPhysB_72,2013_Faribault_PRL_110,
2013_Faribault_PRB_88,2006_Witzel_PRB_74,2006_Yao_PRB_74,2013_Stanek_PRB_88,2014_Uhrig_PRB_90, 
2007_Erlingsson_PRB_70,2008_Coish_PRB_77,2010_Bortz_PRB_81,2010_Coish_PRB_81,2010_Bortz_PRB_82,2012_Barnes_PRL_109,
2014_Hackmann_PRB_89}. Among many decoherence sources, spin-orbit coupling, the hyperfine contact interaction and dipolar interactions between the bath nuclear spins have been widely studied. The simultaneous treatment of all these sources of decoherence is presently beyond reach. In this paper, we focus on the competition between the two latter interactions, neglecting any decoherence due to spin-orbit coupling due to its suppressed influence at low temperatures for localized electrons.

The hyperfine interaction between the electron spin and nuclear spins is an unavoidable source of decoherence in for instance GaAs semiconductor quantum dots.
It is considered to be the dominant source of dephasing at low external fields due to the dynamics of the nuclear spins through the non-secular coupling with the electron spin \cite{2005_Shenvi_PRB_71}. However, at high fields the direct flip-flop processes between electron and nuclear spins are energetically unfavorable due to the energy mismatch between the large Zeeman splitting of the electron spin and the negligible Zeeman splitting of the nuclear spins \cite{Urbaszek_RevModPhys_85} (the g-factor differs by 3 orders of magnitude).  
Only higher order processes of the hyperfine interaction such as electron spin mediated flip flopping will give rise to bath dynamics, which cannot be reversed with for instance spin echo techniques. Thus, at high fields it is not the hyperfine interaction but the dipolar interactions between the nuclear spins that is considered to be the relevant physical process for dephasing \cite{2003_Sousa_PRB_68}. Our primary goal is to take these interactions into account simultaneously and study their mutual competition. 

\begin{figure}[ht]
	\includegraphics[scale=0.25]{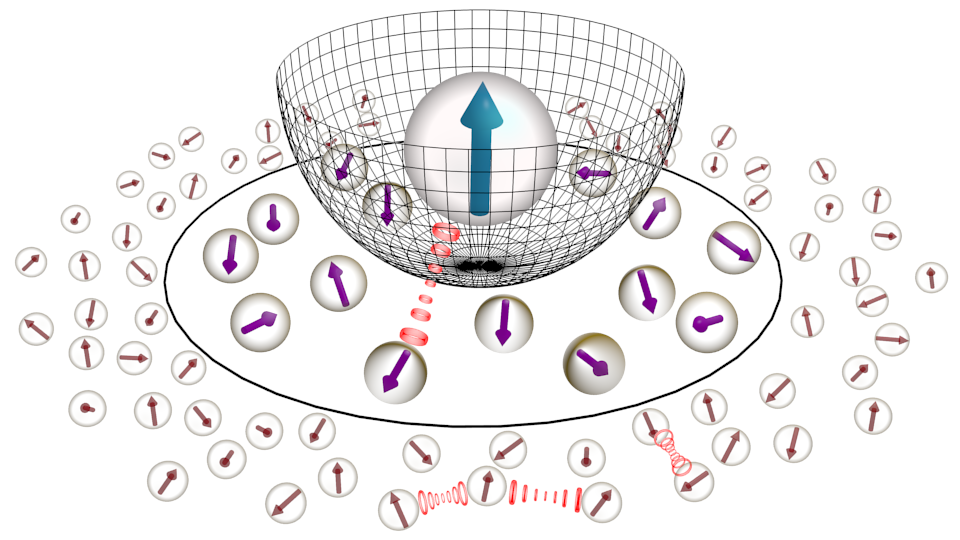}
		\caption{A schematic picture of a semiconductor quantum dot with an electron central spin confined in a harmonic trap. The interactions between the closest nuclear spins and the central spin are described by the full hyperfine interaction including flip-flop terms. For the nuclear spins with hyperfine couplings smaller than or equal to $1/e$, the dipolar interactions among themselves is the dominant contribution to the dynamics. The dynamics of these spins is treated with a mean-field approximation with an effective time-dependent magnetic field in the $z$-direction.}
	\label{fig:quantumdot}
\end{figure}

The interaction between the nuclear spins and the electron spin, which from now on we will call the central spin, is modeled with the full hyperfine contact interaction for those nuclear spins with the strongest coupling. For the more weakly coupled nuclear spins, we consider the dipolar interactions to be the dominant source of dynamics. Following Dobrovitski \textit{et. al} \cite{2009_Dobrovitski_PRL_102}, we model the effect of these interactions on the central spin with a time-correlated Markovian  Gaussian stochastic field in the $z$-direction, also known as an Ornstein-Uhlenbeck (OU) process.  Although it is merely a phenomological mean-field approximation, it has been shown that it is in quantitative agreement with simulations incorporating the actual dipolar couplings for the nuclear spins \cite{2009_Dobrovitski_PRL_102}. More sophisticated methods incorporating microscopic derivations have been used in \cite{2003_Sousa_PRB_68, 2005_Witzel_PRB_72, 2007_Saikin_PRB_75}.

Our secondary goal is to introduce a new integrability-based method applicable to systems which are close to an integrable model, but whose integrability is broken due to the presence of a perturbation.  Possible examples are transverse magnetic fields, or additional interaction terms between two particles. The restrictions on the form of these integrability breaking terms will become clearer in the methods section. The perturbation may furthermore be time-dependent so that driven systems can be studied without additional difficulties. In this paper, as a proof of principle, we consider perturbing the integrable Central spin model with a time-dependent stochastic field on the central spin, representing the effect of dipolar interactions in a mean-field spin bath. 
However, it is important to note that the method described hereafter can be used for any model close to a spin-1/2 integrable model. 

The idea is based on the fact that integrable models in general describe strongly correlated systems and are exactly solvable through the (Algebraic) Bethe Ansatz. The corresponding eigenstates, so-called Bethe states, contain all the information about these correlations, forming a natural starting point when studying integrability-broken models. Integrability thus enables us to treat these interactions more easily as part of a bigger set of interactions. The usefulness of integrability for non-integrable models was already demonstrated in previous works by Delfino \textit{et al.} \cite{1996_Delfino_NPB}, Konik \textit{et. al} \cite{2007_Konik_PRL_98,2009_Konik_PRL_102} and Caux \textit{et. al} \cite{2012_Caux_PRL_109}, and we here wish to extend the settings in which such methods can be used.

The paper is organized as follows: in section \ref{sect:model} the details of the model describing the combination of the two different types of interactions are clarified. In section \ref{sect:methods}, the theoretical approach used to perform the simulations is explained. In the results section the effect of the combination of the hyperfine and dipolar interaction is studied with spin echo (SE) and free induction decay (FID) simulations. Finally, in the last section we discuss and summarize our results.

\section{Perturbing the integrable Central spin model}
\label{sect:model}
The hyperfine contact interaction between the central spin and a finite number $N$ of surrounding nuclear spins is described by the integrable central spin model \cite{1976_Gaudin_JPhys}, with the Hamiltonian given by
\begin{align}
H_{\mathrm{int}} = B_z S_0^z + \sum_{j=1}^N A_j \; \vec{S}_0 \cdot \vec{I}_j,
\label{eq:Hhyperfine}
\end{align}
where $S_0^{\alpha}$ and $I_j^{\alpha}$ are the spin operators of the central spin and the nuclear spins respectively.
All spins are taken to be spin-1/2 particles, and the coupling constants $A_j$ can be chosen freely without destroying the integrability. The influence of the external magnetic field on the nuclear spins is neglected due to the aforementioned 3 orders of magnitude difference between the g-factor of the electron spin and the nuclear spins. 

We consider a coupling distribution corresponding to a Gaussian envelope wave function of a localized electron in 2D  \cite{2004_Coish_PRB_70}
\begin{align}
\psi(r_k) = \psi(0) \exp \left(-\frac{1}{2} \left(\frac{r_k}{l_0}\right)^2\right),
\end{align}
with Bohr radius $l_0$, and the integer $k$ labeling the number of nuclear spins within radius $r_k$ of the central spin. The hyperfine coupling constants are related to the envelope of the wave function through \cite{2003_Schliemann_JPhys}
\begin{align}
A_k = \frac{A}{n_0} | \psi(r_k)|^2,
\end{align}
where $n_0$ is the density of nuclei in the substrate. Given that $k \propto r^2_k$ in 2D, the hyperfine coupling constants yield
\begin{align}
A_k = \frac{A}{N} \exp \left[-\frac{(k-1)}{N_0}\right], \quad \text{with} \quad k=1,...,N
\label{eq:A_hyperfine}
\end{align}
where $N_0$ is the number of nuclear spins within the Bohr radius $l_0$, $N$ is the number of nuclear spins we describe with the hyperfine interaction, and $A$ is the hyperfine coupling strength. 
From here on we will use energy units corresponding to $A/N=1$ and set $N_0$ equal to $N$, such that all $N$ nuclear spins are coupled to the central spin with coupling constants between $1$ and $1/e$ . These bath spins are depicted in Fig. \ref{fig:quantumdot} as the spins closest to the central spin. Because of the strong coupling with the central spin, the dipole-dipole interactions among these bath spins are considered to be of minor importance as compared to the hyperfine interaction, and as such are neglected.

The effect of the dynamics of the more weakly coupled bath spins, indicated in Fig. \ref{fig:quantumdot} as the smallest nuclear spins, is described using a mean-field treatment. It is modeled as an OU process \cite{2009_Dobrovitski_PRL_102}, with time-dependent correlations $\braket{B(t)B(0)} = b^2 \exp(-Rt)$, where the dispersion $b$ is determined by the fluctuations of the Overhauser field of the bath spins involved 
\footnote{The expression for the dispersion $b$ is obtained by setting $N_0=N$. In the case where one wants to look at the limits for $N\rightarrow\infty$ and $N_0\rightarrow \infty$ the geometric series only converges when the limit for $N\rightarrow\infty$ is taken before the limit $N_0\rightarrow \infty$.}

\begin{align}
b = \sqrt{\sum_{k=N+1}^{\infty} A_k^2}  = e^{-1} \sqrt{\frac{1}{1-e^{-2/N}}}.
\label{eq:OU_b}
\end{align}
Here, $R$ denotes the correlation decay rate of the dynamics of the mean-field spin bath, and quantifies the strength of the dipolar interactions within the bath. The speed of the bath is determined by comparing $R$ to the dispersion $b$. The mean value of the fluctuating field $B(t)$ is set to zero unless indicated otherwise, mimicking a spin bath with on average zero polarization. We furthermore only take into account the effect on the $z$-component of the central spin and neglect the effects on its transverse components. The mean-field treatment of the dipole-dipole interactions thus yields an extra term $V(t) = B(t) S_0^z$ in addition to the Hamiltonian given in eq. \eqref{eq:Hhyperfine}.

The effective fluctuating field causes the Zeeman splitting of the central spin to diffuse over time, an effect also called spectral diffusion. Except for dipole-dipole interactions within the nuclear spin bath, other processes can also lie at the origin of spectral diffusion, such as small fluctuations of the external magnetic field in experiments. Nevertheless, we only focus on the dipole-dipole interaction within the bath.

Let us now briefly comment on some of the approximations that the model is based on. First, we neglect the effect of the back-action of the central spin on the mean-field bath fluctuations. Although this is correct to first order, it would be interesting to see how our results are influenced by using a more sophisticated model for dipolar interactions. However, this is beyond the scope of this paper.

Second, we neglect the direct interaction between the strongly coupled nuclear spins and the mean-field bath. Nevertheless, the influence of the mean-field bath is indirectly felt by the closer nuclear spins through the hyperfine interaction with the central spin. 
As will be discussed in the methods section, we will use the eigenstates of the hyperfine Hamiltonian as the computational basis. This has the advantage of taking into account the interaction with the central spin without any approximation, thus also representing the indirect influence of the fluctuating field on the nearby nuclear spins in a non-perturbative way.

In the next section we will describe the method used in order to study both the hyperfine interaction as well as the effective fluctuating magnetic field. 

\section{Methods}
\label{sect:methods}
Let us consider a system consisting of an integrable part and a perturbation which breaks integrability
\begin{align}
H = H_{\mathrm{int}} + V,
\end{align}
thus preventing us from solving for the eigenstates using the Bethe Ansatz. For these types of situations the method described below provides a way to profit from 
the fact that the model is close to an integrable one. It is based on the idea of using the eigenstates of the integrable part of the model $H_{\mathrm{int}}$ as basis states for the time evolution operator of the integrability-broken Hamiltonian. By splitting up the time evolution operator into small time steps we can time evolve a non-integrable system using the integrable eigenstates.

In this paper we consider the Gaudin central spin model of eq. \eqref{eq:Hhyperfine} to be our integrable model. Its eigenstates are of the form
\begin{align}
\ket{\{\lambda_1,\lambda_2,...,\lambda_M\}} = \prod_{k=1}^M S^+(\lambda_k)\ket{\Downarrow; \downarrow\downarrow ... \downarrow},
\end{align}
with generalized raising operators
\begin{align}
S^+(\lambda) = \frac{S^+_0}{\lambda - \epsilon_0} + \sum_{j=1}^N \frac{I^+_j}{\lambda - \epsilon_j}.
\end{align}
The rapidities $\{\lambda_i\}$ that determine the eigenstates can be obtained by solving the Richardson-Gaudin equations \cite{2011_Faribault_PRB_23,GaudinBOOK,*GaudinBOOK_translation}
\begin{align}
-2\beta - \sum_{j=1}^{N+1} \frac{1}{\lambda_i - \epsilon_j} + 2\sum_{k \neq i} \frac{1}{\lambda_i - \lambda_k} = 0,
\label{eq:RGeq}
\end{align}
with $\beta = -B_z/2$ and $1/(\epsilon_0-\epsilon_j)=A_j$.
The corresponding eigenenergies are given by 
\begin{align}
E\left(\{\lambda_i\}\right) = \beta + \frac{1}{2} \sum_{j\neq 0} \frac{1}{\epsilon_0 - \epsilon_j} + \sum_{l=1}^{M} \frac{1}{\lambda_l - \epsilon_0}.
\end{align}
As explained in section \ref{sect:model} we perturb the Gaudin central spin model with a time-dependent field on the central spin $V(t) = B(t)S^z_0$.
The next step is to express the time evolution operator in the integrable basis.

\subsection{Approximating the time-evolution operator}
In order to time evolve an initial state under the influence of the integrabilty-broken Hamiltonian, we split up the time-evolution operator into small time steps, and approximate it with a second-order Suzuki-Trotter decompostion  \cite{2005_Suzuki_LectureNotes}. For the case of a time-dependent perturbation this yields
\begin{align}
U(t,t+dt) \approx U^{(2)}= e^{-i\frac{dt}{2} H_{\mathrm{int}}} e^{-i dt V\left(t+\frac{dt}{2}\right)} e^{-i\frac{dt}{2}  H_{\mathrm{int}}},
\label{eq:S2}
\end{align}
with an error of the order $dt^3$.
Using the eigenstates $\ket{\psi_i}$ of the spin-1/2 integrable model with corresponding eigenvalues $E_i$ as the computational basis, eq. \eqref{eq:S2} yields
\begin{align}
\braket{\psi_i | U^{(2)} | \psi_j} = e^{-i \frac{dt}{2} \left(E_i + E_j\right)} \braket{\psi_i | e^{-i dt V\left(t+\frac{dt}{2}\right)} | \psi_j},
\label{eq:S2_bethe}
\end{align}
where $U^{(2)}$ depends on the time step $dt$ and in the case of a time-dependent perturbation also on time $t$.

The expression for the exponent of an operator such as the one present in eq. \eqref{eq:S2_bethe} is in general not known from the Algebraic Bethe Ansatz.
However, in the following subsection, we show how to obtain the required matrix elements in the basis of Bethe states.

\subsection{Obtaining unknown local operators in the Bethe basis}
\label{section:loc_op}

Suppose we have a spin-1/2 integrable model with $N$ particles, and consider a local operator $\hat{O_i}$ acting on some particle $i$. Just like any wave function of a spin-1/2 model, a Bethe wave function can be written as
\begin{align}
\ket{\psi}_M = \overbrace{\sum_{k} c^+_k \ket{\uparrow}_{\mathrm{i}} \otimes \ket{\phi_k}_{M-1} }^{\ket{\psi^+}_{M}}  + \overbrace{\sum_{k} c^-_k \ket{\downarrow}_{\mathrm{i}} \otimes \ket{\phi_k}_{M}}^{\ket{\psi^-}_M},
\label{eq=decompBethe}
\end{align}
where $\{\ket{\phi_k}\}$ is a complete orthonormal set of states of the spin-1/2 particles unaffected by $\hat{O_i}$. Furthermore, $c^+_k$ and  $c^-_k$ are coefficients such that $\sum_k |c^+_k|^2 + |c^-_k|^2 = 1$. The index $M$ labels the sector of spin flips contained in $\ket{\phi_k}_M$, such that $\tensor*[_M]{\braket{\phi_k|\phi_l}}{_{M^{\prime}}} = \delta_{k,l} \delta_{M,M^{\prime}}$. 
Using eq. \eqref{eq=decompBethe}, the nonzero matrix elements of $\hat{O_i}$ can then be written as
\begin{align}
\tensor*[_M]{\braket{\psi | \hat{O}_i |\tilde{\psi}}}{_M} 
=& \tensor[_M]{\braket{\psi^+| \hat{O}_i |\tilde{\psi}^+}}{_M} 
+  \tensor[_M]{\braket{\psi^-|\hat{O}_i |\tilde{\psi}^-}}{_M} \notag \\
\tensor*[_{M}]{\braket{\psi | \hat{O}_i |\tilde{\psi}}}{_{M-1}}  
=& \tensor[_M]{\braket{\psi^+| \hat{O}_i |\tilde{\psi}^-}}{_{M-1}} \notag \\
\tensor*[_{M-1}]{\braket{\psi | \hat{O}_i |\tilde{\psi}}}{_{M}}  
=& \tensor[_{M-1}]{\braket{\psi^-| \hat{O}_i |\tilde{\psi}^+}}{_{M}}
\end{align}
where the third case is trivially related to the second case for a hermitian operator $\hat{O}$. All other matrix elements are zero due to a mismatch in excitation sectors $M$ and $M^{\prime}$ of the states describing the unaffected spins. 

Using the explicit form of eq. \eqref{eq=decompBethe} for $\ket{\psi^+}$ and $\ket{\psi^-}$ we obtain
\begin{align}
\label{eq=plusplus}
\tensor[_M]{\braket{\psi^+|\hat{O}_i|\tilde{\psi}^+}}{_M}
=&\sum_{m,l} (c^+_m)^* \tilde{c}^+_l  \tensor*[_{M-1}]{\braket{\phi_m|\phi_l}}{_{M-1}}
\braket{\uparrow|\hat{O}_i |\uparrow} \notag \\
= & \hat{O}_i^{\uparrow \uparrow}  \; \sum_{l} (c^+_l)^* \tilde{c}^+_l  \notag \\
=& \hat{O}_i^{\uparrow \uparrow}  \; \tensor*[_M]{\braket{\psi|S^+_i S^-_i|\tilde{\psi}}}{_M}  \\
\label{eq=minmin}
 \tensor[_M]{\braket{\psi^-|\hat{O}_i|\tilde{\psi}^-}}{_M} 
= & \hat{O}_i^{\downarrow \downarrow} \; \sum_{l} (c^-_l)^* \tilde{c}^-_l  \notag  \\
=& \hat{O}_i^{\downarrow \downarrow} \;\tensor*[_M]{\braket{\psi|S^-_i S^+_i|\tilde{\psi}}}{_M}  \\
\label{eq=plusmin}
\tensor[_M]{\braket{\psi^+|\hat{O}_i|\tilde{\psi}^-}}{_{M-1}}
= & \hat{O}_i^{\uparrow \downarrow} \; \sum_{l} (c^+_l)^* \tilde{c}^-_l  \notag \\
=&  \hat{O}_i^{\uparrow \downarrow} \;  \tensor*[_M]{\braket{\psi|S^+_i|\tilde{\psi}}}{_{M-1}},
\end{align}
where, $\hat{O}_i^{\uparrow \uparrow},\hat{O}_i^{\uparrow \downarrow},... $ are the matrix elements of $\hat{O}_i$ in the basis of a single spin-1/2 particle.
Furthermore, note that the operators $S^+_iS^-_i$ and $S^-_iS^+_i$  are equivalent to $\frac{1}{2} + S^z_i$ and $\frac{1}{2} - S^z_i$ respectively in spin-1/2 models. the matrix elements of these operators, as well as the $S^+_i$ operator, can be obtained through the application of Slavnov's formula for the scalar product between on-shell and off-shell Bethe states \cite{1989_Slavnov_TMP_79_English}, and are represented by single determinants \cite{2003_Links_JPhysA}. Thus, every matrix element of a local operator $\hat{O}_i$ acting on one of the spin-1/2 particles of an integrable model, can be computed in a numerically efficient way.

A straightforward generalization can be made for the case of an operator that acts on two spin-1/2 particles, such as an interaction term. The restrictions on the operator $\hat{O}$ are such that it can only affect a few number of particles, since the decomposition for Bethe states otherwise becomes computationally too demanding.

The decompositions of eqs. (\ref{eq=plusplus},\ref{eq=minmin},\ref{eq=plusmin}) can be used to express the exponent of the perturbation $V$ in eq. \eqref{eq:S2_bethe} in the Bethe basis, useful for cases where analytical expressions are not available. This gives us access to the second-order Suzuki-Trotter approximation of the time-evolution operator for a finite timestep $dt$. In this paper we use timesteps of typical size $dt = 0.005$, leading to an error of $6.25 \times 10^{-4}$ for a time evolution up to $t = 25$ ($5000$ timesteps). Furthermore, the timestep is always in the regime $dt \ll 1/R$, such it is smaller than the time scale for the change in the time-dependent magnetic field.

We thus study the time evolution of a non-integrable model, using an integrable set of states as the computational basis. The method just described is only based on the general form of the wavefunction of a spin-1/2 model, and the assumption that part of the Hamiltonian is integrable. It can thus be applied to all spin-1/2 integrable models with a local perturbation, such as Hamiltonians close to the integrable XXZ chain.

\subsection{Comparison to other numerical methods}
Before proceeding to the main results of our paper, we briefly want to shed light on the comparison of the proposed method with other well-known related numerical schemes for the time evolution of an initial state.

When comparing with exact diagonalization (ED), the obvious difference is that our method can handle time-dependent perturbations, something that is not the case for ED.
Other numerical schemes such as the L\'anczos procedure (\cite{2005_Manmana_AIPConf}), which is an exact diagonalization method based on the Krylov space expansion of the time-evolution operator for short time steps $dt$, are also capable of handling such perturbations. However, two important differences should be highlighted.
First, the error introduced in our representation of the time-evolution operator in the Bethe eigenstates resides solely in the Suzuki-Trotter approximation we use to split up the evolution under the action of the integrable part and the perturbation. The use of the Bethe eigenbasis allows us to represent the exponent of the integrable part of the Hamiltonian in a trivial way, and we have shown that we can compute the matrix elements of the exponent of the perturbation in an exact manner. The estimate of the order of magnitude of the error introduced by the Suzuki-Trotter approximation is thus well controlled, and only determined by the time step $dt$. On the contrary, for the case of  Krylov space based methods, the error not only depends on the time step $dt$, but also on the spectrum of the Hamiltonian \cite{2005_Manmana_AIPConf}.

Second, the number of matrix-vector multiplications required to perform the evolution of the wavefunction over a time step $dt$ is different for these two methods. Where the L\'anczos procedure requires multiple matrix-vector multiplications in order to generate all the Krylov vectors, our methods only requires one matrix-vector per time step. It is important to note that this is precisely due to the combination of the use of the Bethe eigenstates as the computational basis, and the Suzuki-Trotter approximation of the time-evolution operator. If one would perform a Suzuki-Trotter approximation using a local product basis, multiple matrix-vectors operations would still need to be performed in order to capture the time evolution due to the integrable part of the Hamiltonian.

Compared to the Hamiltonian representation in the local product basis, which is used in the matrix-vector operations for the Krylov vectors, the operator in eq. \eqref{eq:S2_bethe} is less sparse. However, it only couples states in the same sector, or neighbouring sectors, thus maintaining a sparse structure. We believe that the numerical performance of the proposed method and Krylov vector methods is similar for the case of central spin models.


Finally, let us compare our method to the well-known density matrix renormalization group method (DMRG), which was recently introduced for the central spin model in \cite{2013_Stanek_PRB_88}. When it comes to the number of particles accessible through both methods, DMRG greatly exceeds the capabilities of the proposed method in this paper. With t-DMRG, time evolution can also be studied, albeit limited to short times due to the truncation in the Hilbert space that is used in DMRG. However, we believe our method is more flexible when it comes to the types of interactions that can be taken into account. Adding an interaction between two nuclear spins as a perturbation to the central spin model would not pose any additional difficulties for our method. To our knowledge, this would complicate the numerical scheme for DMRG.

\section{Results}
\label{sect:results}
\begin{figure*}[htp]
    \begin{center}
	\includegraphics[scale=1.]{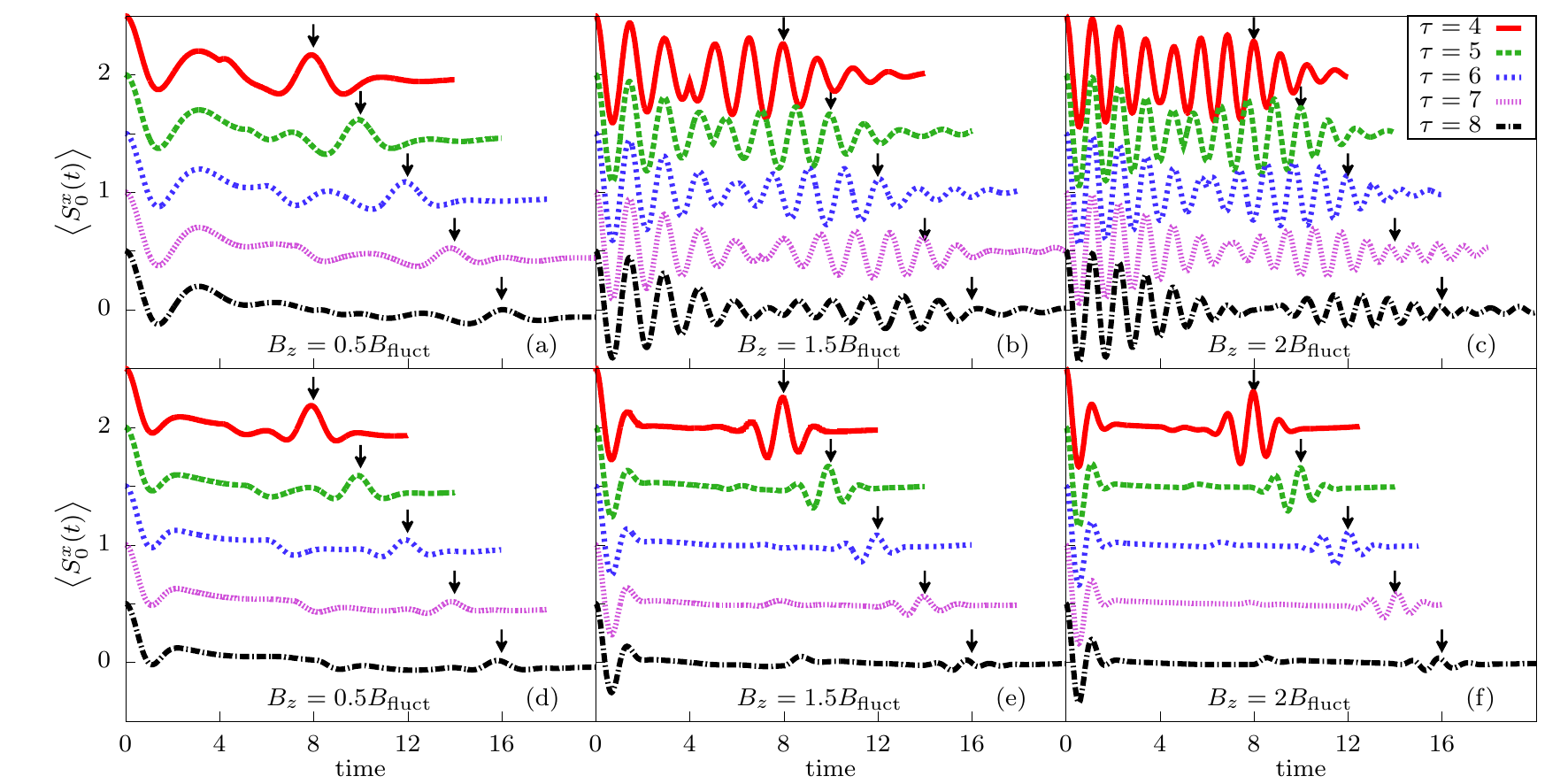}
		\caption{The expectation value of $S^x_0$ is plotted for different pulse times $\tau$ for $N=14$ nuclear spins with couplings between $1$ and $1/e$. In (a)-(c) the closest nuclear spins with hyperfine couplings in the range $\left[1,1/e\right)$ are initially in a N\'eel-like pure state. The magnetic fields are given by $B_z = 0.5 B_{\mathrm{fluct}}$ in (a), $B_z = 1.5 B_{\mathrm{fluct}}$ in (b), and $B_z = 2 B_{\mathrm{fluct}}$ in (c). On the contrary, in (d)-(f) the closest nuclear spins are initially in a random initial state. The same magnetic fields as in (a)-(c) are considered. All curves are offset from one another by $1/2$ for clarity and are averaged over $100$ OU processes to mimic the dynamics of the mean-field bath. Furthermore, in each run in (d)-(f) a different random initial state was generated. The time step used for the Suzuki-Trotter decomposition is $dt = 0.005$, the decay rate of the fluctuating field is fixed to $R = 0.01$ and the dispersion $b$ is given by eq. \eqref{eq:OU_b} and has an approximate value of $b \approx 1.008$. The arrows indicate the expected arrival time of the echo at $t = 2 \tau$. For magnetic fields $B_z = 1.5B_{\mathrm{fluct}}$ and $B_z = 2B_{\mathrm{fluct}}$ the echo signal is pushed down in an asymmetric fashion. This leads to a shift in the echo for (b) and (c), but no visible shift for (e) and (f). For $B_z = 0.5B_{\mathrm{fluct}}$ there is no clear distinction between the plots for the two initial states and no shift present.}
	\label{fig:echoeshift}
	      \end{center}
\end{figure*}

Using the integrability-based method described above, we consider the example of a semiconductor quantum dot model including both hyperfine interactions and a mean-field treatment of dipolar interactions. The combination of these interactions gives us two dynamical sources of decoherence. The first is given by the dynamics of the closest nuclear spins through the hyperfine interaction, either by direct flip-flopping with the central spin, or by hyperfine mediated exchange processes through the central spin \cite{2009_Cywinski_PRL_102,2009_Cywinski_PRB_79}. The latter process is dominant at high fields for which it can be estimated to have an effective interaction strength proportional to \cite{2005_Shenvi_PRB_71}
\begin{align}
\propto \frac{\sum _{j,k} A_j A_k}{B_z}. 
\label{eq:hyp_exchange}
\end{align}  
The energy cost associated to this process is proportional to $A_j-A_k$, such that two nuclear spins with comparable hyperfine couplings can flip-flop even at high fields. 
The second dynamical source of decoherence is given by the spectral diffusion modeling the dynamics of the mean-field bath with coupling constants lower than $1/e$. 

Moreover, static sources such as different nuclear environments for different quantum dots also contribute to decoherence of the quantum dot, in particular to the ensemble dephasing time  $T_2^*$. However, in 1950 Hahn \cite{1950_Hahn_PR_80} proposed a spin echo pulse sequence which removes the decoherence due to these static sources. The decay of the spin echo sequence is then given by the intrinsic dephasing time $T_2$. The effect of hyperfine couplings in spin echo simulations has been studied previously in \cite{2005_Shenvi_PRB_71,2009_Cywinski_PRL_102,2009_Cywinski_PRB_79,2006_Yao_PRB_74,2010_Cywinski_PRB_82} and spin echo simulations focusing on the effect of dipole-dipole interactions have been performed among others by  \cite{2003_Sousa_PRB_67,2003_Sousa_PRB_68,2005_Witzel_PRB_72,2006_Witzel_PRB_74,2007_Witzel_PRB_76,2006_Yao_PRB_74,2007_Saikin_PRB_75,
2010_Witzel_PRL_105}. In order to study the competition between these two interactions we have performed spin echo simulations for varying magnetic fields $B_z$ and bath correlation decay rates $R$, and compared these for different initial states.

\subsection{Spin echo simulations}

Although a standard spin echo protocol consists of a $\pi_y/2-\tau-\pi_x-\tau-\pi_y/2$ pulse sequence, with free precession time $\tau$, we assume all pulses to be ideal, and focus on the $\tau-\pi_x-\tau$ part where decoherence through the interaction with the nuclear spins during the free precession time will play an important role. The operator representing the $\pi$-pulse around the $x$-axis acting on the central spin is given by
$(S_0^+ + S^-_0)$.

We consider two different types of initial bath configurations for the nuclear spins with hyperfine couplings between $1$ and $1/e$. The first corresponds to a pure state of N\'eel type: $\ket{\psi_0} =\frac{1}{\sqrt{2}}\left( \ket{\Uparrow} + \ket{\Downarrow}\right) \otimes \ket{\uparrow \downarrow \uparrow \downarrow  \uparrow \downarrow  ...}$, where the ordering of the bath spins is such that the strongest coupled nuclear spin is the first. The second initial state is given by a more experimentally realistic state, $\ket{\psi_0} =  \frac{1}{\sqrt{2^{N+1}}}\left( \ket{\Uparrow} + \ket{\Downarrow}\right) \otimes_{k=1}^N \left(\ket{\uparrow} + e^{i\phi_k}\ket{\downarrow}\right)$, where $\phi_k$ are random phases. It was shown by Schliemann \textit{et al.} \cite{2002_Schliemann_PRB_66} that these random states mimic a mixed initial bath configuration, relevant for studying infinite temperature baths. The overlap of these initial states with the Bethe states, as well as the matrix elements required for the $\pi$-pulse can be obtained from the aforementioned Slavnov determinants \cite{1989_Slavnov_TMP_79_English,2003_Links_JPhysA}.

We first study the time evolution of the expectation value $\braket{S^x_0}$ (the real part of $\braket{S^+_0}$) for various pulse times 
$\tau$ of the echo sequence and the two initial states. For clarity, the calculation proceeds in the following way: the initial state is evolved in time with the Suzuki-Trotter approximation given by eq. \eqref{eq:S2_bethe}, yielding $\ket{\psi(t)}$ as a linear superposition of Bethe eigenstates. The expectation value $\braket{S^+_0(t)}$ is then readily obtained from the matrix elements of $S^+_0$ in the Bethe basis.

We consider $N=14$ bath spins treated with the integrable hyperfine interaction with coupling constants between $1$ and $1/e$, and treat the effect of the remaining bath spins with a time dependent stochastic field described by slow bath dynamics such that $R=0.01$. 

In a previous study \cite{2005_Shenvi_PRB_71} it was shown that the decay of the spin echo signal within a central spin model with only hyperfine interaction is governed by the Overhauser field due to spin fluctuations 
\begin{align*}
B_{\mathrm{fluct}} = \sqrt{\sum_{k=1}^N A_k^2}.
\end{align*}
For external fields larger than this critical field it was reported that a large portion of the original value of the spin can be recovered using a spin echo sequence. This can be explained by the fact that the Overhauser field fluctuations cannot overcome the Zeeman gap caused by the external field, thus making direct flip-flopping between nuclear spins and the central spin a negligible process.

In order to study the role of this critical field in the presence of an additional fluctuating bath, we have performed spin echo simulations for values of the external field below and above the Overhauser field fluctuations. The results for $B_z = 0.5 B_{\mathrm{fluct}},1.5 B_{\mathrm{fluct}}, 2 B_{\mathrm{fluct}}$ are shown in Fig. \ref{fig:echoeshift}.

For external fields larger than $B_{\mathrm{fluct}}$ the shape of the echoes for the initial pure state differs significantly from the echoes for the random initial state. The echoes for the N\'eel state show an apparent early arrival of the echo as compared to the expected arrival time at $2\tau$, an effect which is not observed in the case of the random initial bath configuration. 

This shift can be attributed to an asymmetrical decay of the echo due to the dynamics of the mean-field bath through the dipolar interactions. This source of dynamics in the bath cannot be reversed with a spin echo sequence, thus leading to an additional decay of the spin echo on top of the decay due to the transverse hyperfine interactions.
The asymmetry of the decay is due to a larger effect of decoherence by the fluctuating field on the later parts of the echo signal as compared to the earlier parts. 

The reason why this asymmetry only leads to a visible shift for the pure state, is the difference in dephasing times, which is much longer in the case of the N\'eel-like state \cite{2011_Faribault_PRB_23}. This leads to an echo signal which is much wider for the pure state, such that the asymmetry will be stronger, leading to a shift in the maximum of the echo. Nevertheless, it should be noted that even though there is no visible shift in the echoes for the random initial state, the echoes are asymmetric around $t=2\tau$.

Another noteworthy feature is that the shifts of the echoes at $B_z = 1.5 B_{\mathrm{fluct}}$ and $B_z = 2 B_{\mathrm{fluct}}$ seem to be of the same size, which indicates that it is the fluctuating field that is responsible for the shift, and not the dynamics of the closer nuclear spins through the hyperfine mediated exchange for which the effective interaction parameters are field dependent (see eq. \ref{eq:hyp_exchange}).

In contrast to the cases with high magnetic field, there is no visible shift in the echo for either initial states when $B_z = 0.5 B_{\mathrm{fluct}}$. This implies that even in the presence of a fluctuating bath due to dipolar couplings, the dynamics of the nuclear spins through the hyperfine coupling is the dominant source of dephasing at short times, and is competing at longer times with the mean-field bath fluctuations, not leaving enough room for the asymmetry of the echo to lead to a shift.

\begin{figure}[htp]
    \begin{center}
	\includegraphics[scale=1.]{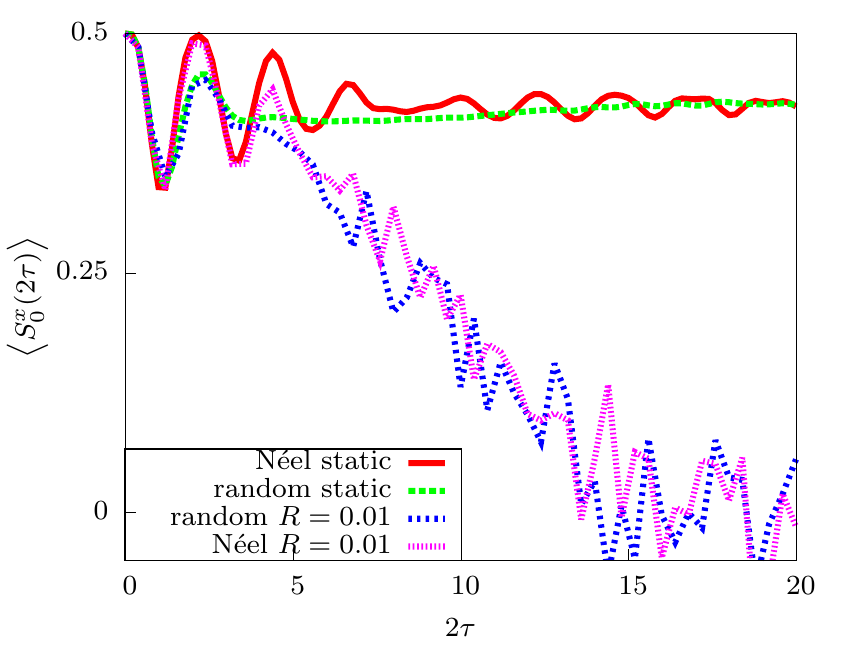}
		\caption{For $N=14$ and $B_z = 2B_{\mathrm{fluct}}$, the expectation value  $\braket{S_0^x(2\tau)}$ is plotted for the N\'eel pure initial state and an average over initial random states. The red and green curves show the results for the integrable model with only hyperfine interactions, corresponding to a static mean-field bath with zero average polarization. The purple and blue curves show the echo envelope for a slow mean-field bath with $R=0.01$, averaged over 40 OU processes for each point (every point thus requires 40 time evolution computations of which only the value $\braket{S_0^x(2\tau)}$ is saved).} 
	\label{fig:envelopecompneelrand} 
	      \end{center}
\end{figure}

Even though the shape of the echo signal differs significantly between the two initial states for fields larger than $B_{\mathrm{fluct}}$, the actual value of the signal at the expected arrival time $2\tau$ does not show a strong dependence on the initial state for $2\tau > 5$, as can be seen in Fig. \ref{fig:envelopecompneelrand}.
The expectation value  $\braket{S_0^x(2\tau)}$, also referred to as the echo envelope, is shown as a function of the total free precession time $2\tau$ for the two different initial states, where the corresponding cases for static mean-field baths are also shown. For short times the hyperfine flip-flopping terms are the dominant source of decoherence, such that the curves with the slowly varying mean-field bath follow their corresponding curves for the static mean-field bath closely.
For longer times, the dynamics of the mean-field spin bath becomes the primary source of decoherence, and since these dynamics are independent of the initial state, the decay for the two initial states of the echo envelope becomes of the same order of magnitude.

\begin{figure}[htp]
    \begin{center}
	\includegraphics[scale=1.]{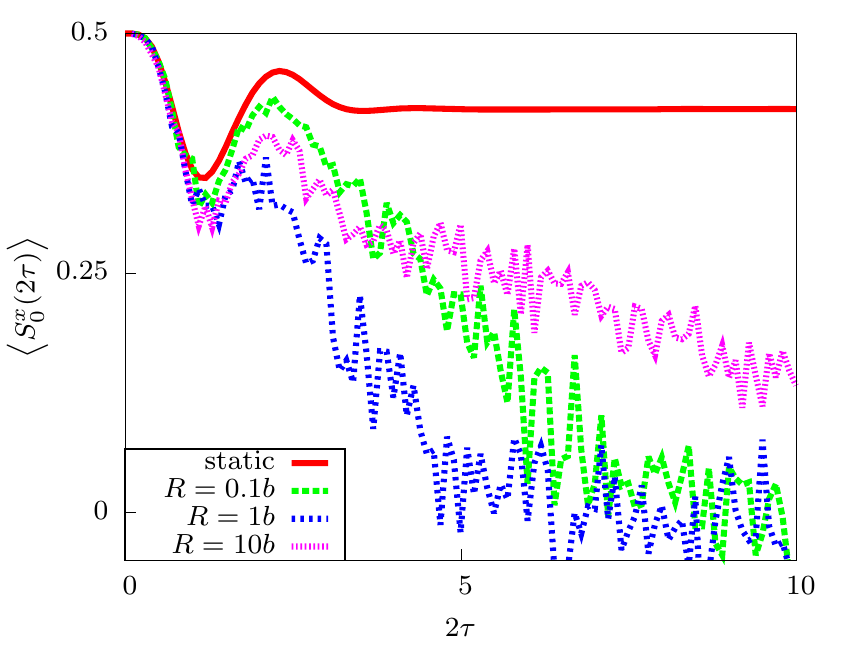}
		\caption{Echo envelopes for $N=14$ nuclear spins treated with the full hyperfine coupling, an external field of $B_z = 2 B_{\mathrm{fluct}}$ and varying mean-field bath correlation decay rates $R$. All curves corresponding to  a non static mean-field bath have been computed by averaging over $50$  OU processes. The mean-field bath fluctuations are most destructive for $R = b$, and show a slower decay for $R = 0.1b$ and $R = 10b$.} 
		\label{fig:echodiffR}
	      \end{center}
\end{figure}

Next, we will consider the effect of the correlation decay rate $R$ of the mean-field bath on the spin echo envelope. We only show results for a magnetic field $B_z = 2B_{\mathrm{fluct}}$, since the effect of dipole-dipole interactions is believed to be the dominant source of decoherence for large external fields. We furthermore focus on the random initial bath configuration as it is the most realistic initial state. The spin echo envelopes are shown in Fig. \ref{fig:echodiffR} together with the result for the integrable case for which the effect of the mean-field bath has been neglected. 

Similar to what was observed in \cite{2009_Dobrovitski_PRL_102} for the case of Rabi oscillations, we find that the decay of the echo envelope due to the mean-field bath changes nonmonotonically with $R$. It is strongest for $R \sim b$, and becomes less effective for $R < b$ and $R > b$. Fig. \ref{fig:echodiffR} furthermore shows that the slow bath with $R = 0.1b$ causes a faster decay than the bath with $R = 10b$, for which we enter the motional narrowing regime where the dynamics of the mean-field bath is faster than the dynamics of the central spin. When $R \gg b$,  the effect of the mean-field bath on the central spin is effectively averaged out, leading to an echo envelope curve which approaches the result of the integrable static case.

An intuitive picture describing our results is given by the following reasoning. The quantity $R$ represents the rate of change of the Larmor frequency of the additional stochastic field on the central spin. The typical Larmor frequency is given by the standard deviation of the Gaussian distribution, namely $b$. 
If the change in the stochastic field is much faster than the typical Larmor frequency ($R\gg b$), no Larmor precession due to the stochastic field will have a noticeable influence (motional narrowing). 

If, in the other limit the change in the stochastic field is much slower than the rate of the typical larmor frequency, the central spin will have performed many Larmor precessions before feeling a noticeable difference in Larmor frequency. 
The dynamics is then in the adiabatic regime where the evolution is described by the eigenstates of the slowly changing integrable model. Since the change is so slow this is only noticeable on longer times, thus leading to a small decay due to the change in magnetic field.

In between these two limiting cases, there must be an ``optimal'' decay rate, such that a considerable portion of one Larmor precession corresponding to a given value of the stochastic field can be traced by the central spin, before the Larmor frequency has changed significantly. Naturally, the portion of Larmor precession that the central spin traces should not be much larger than the typical Larmor frequency, otherwise leading to an effectively static field over multiple Larmor precessions. The rate with which the stochastic field decorrelates to $1/e$ should thus be of the order of magnitude of $b$. Our results are in agreement with this order of magnitude estimation.

\subsection{Zero Field Free Induction Decay}
\begin{figure}[htp]
    \begin{center}
	\includegraphics[scale=1.]{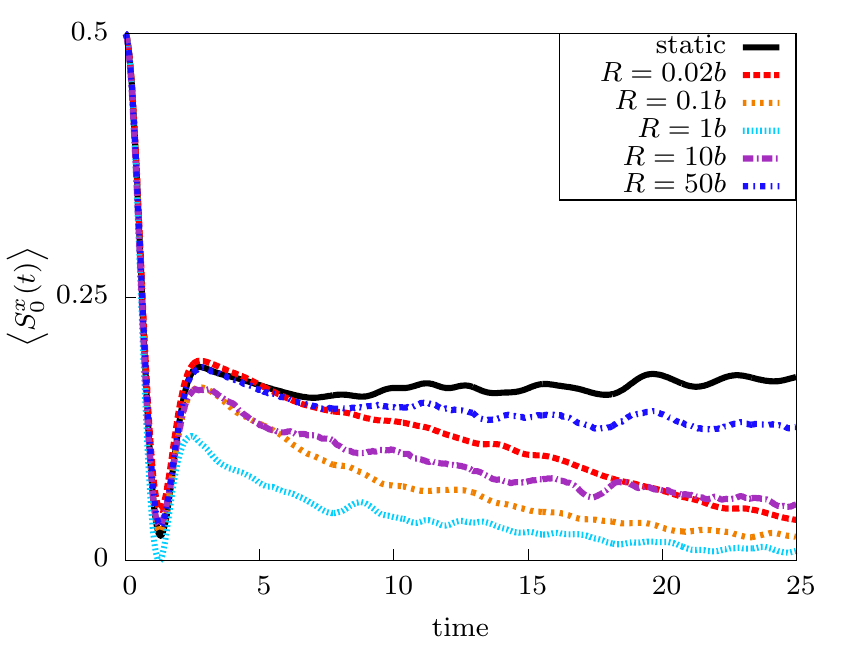}
		\caption{The free induction decay of $\braket{S^x_0}$ for different mean-field bath speeds $R$. The results have been averaged over $100$ OU processes, with a newly generated initial random state for each run. For comparison the integrable case is shown as well. Similar to the case of spin echo simulations, the mean-field bath with $R = b$ shows the fastest decay of the correlation function. } 
	\label{fig:zerofieldrandom}
	      \end{center}
\end{figure}

Another interesting feature of the integrable central spin model emerges at zero external field. Both semiclassical and full quantum treatments \cite{2002_Merkulov_PRB_65,2013_Faribault_PRB_88,2014_Uhrig_PRB_90} of the hyperfine interaction have predicted a non-decaying finite fraction of the initial value of the central spin. Our last results focus on the effect of dipolar interactions in a mean-field bath on this non-decaying fraction. Since the Richardson-Gaudin equations in \eqref{eq:RGeq} cannot be solved for $B_z = 0$, we solve them for a small field and set the average of the OU process such that it cancels the magnetic field, bringing us to the effective $B_z=0$ case.
The results are shown in Fig. \ref{fig:zerofieldrandom} for an average over random initial states and varying bath speeds.

As can be expected, the dynamics of the mean-field bath destroys the non-decaying fraction $\braket{S^x_0}$ caused by the closer nuclear spins.
Similar to the spin echo simulations, the curve corresponding to the mean-field bath with correlation decay rate $R=b$ shows the fastest decay, again in agreement with the intuitive picture described above. Furthermore, the cases for $R = 10b$ and $R = 0.1b$ follow each other closely for short times, after which the slower bath leads to faster decay. The curves shown for $R=50b$ and $R=0.02b$ reflect the same behaviour, with the fast bath ($R = 50b$) approaching the static curve.

\section{Conclusion}
Summarizing, we have proposed a new integrability-based method to treat time-dependent integrability-breaking perturbations, and applied it to the central spin model describing semiconductor quantum dots. Using the integrable basis we treat the hyperfine contact interaction between the central spin and the closest nuclear spins without any approximation. The influence of the dipolar interactions between the nuclear spins more weakly coupled to the central spin is taken into account using a mean-field treatment and represented by a stochastic field in the $z$-direction.

We have shown that the hyperfine contact interaction and the dipole-dipole interaction represented by the mean-field treatment compete in different ways in spin echo simulations, depending on the size of the external magnetic field. For external fields lower than the fluctuations of the hyperfine Overhauser field, the hyperfine interaction is the dominant decoherence source at short times. At longer times it competes with the mean-field fluctuations, leading to a narrow echo that decays due to the mean-field bath fluctuations. 

Our main result is the observation of an asymmetry in the echo at high fields, present for both the N\'eel-like initial state as well as the more realistic random bath configuration. We argue that this asymmetry can be attributed to the fact that later parts of the echo signal have suffered more decoherence caused by the mean-field bath fluctuations compared to the earlier parts. Moreover, due to the wider echo signal of the N\'eel state (larger $T_2^*$), the asymmetry gives rise to a shift in the echo maximum.

The results for the zero field free induction decay show that the non-decaying fraction of $\braket{S^x_0}$, which is present in the integrable Gaudin central spin model, is destroyed at longer times by mean-field bath fluctuations simulating the dipole-dipole interactions.
Furthermore, both the spin echo simulations as well as the free induction decay simulations show that the decay of $\braket{S^x_0}$  exhibits a non-monotonic dependence on the decay rate $R$ of the stochastic field, where $R \approx b$ leads to the strongest decay.

A possible improvement to our work could be to increase the number of nuclear spins with hyperfine couplings between $1$ and $1/e$, which have been treated in an exact manner.
However, free induction decay simulations at zero field in \cite{2013_Faribault_PRB_88} showed that for the particular coupling distribution of eq. \eqref{eq:A_hyperfine}, the value of the non-decaying fraction of $\braket{S^x_0}$ shows a very weak dependence on system size $N$. We thus do not expect a significant change in behaviour for the zero field case when taking into account more nuclear spins with couplings between $1$ and $1/e$. This leads us to believe that a reduced discretization of couplings for the spin echo simulations will not influence the results too strongly either.

It should furthermore be noted that the number of particles in this example is limited by the fact that every curve has to be averaged over different realizations of the Ornstein-Uhlenbeck process in order to get reliable results. If this method would be applied to a case where the perturbation is not of a stochastic nature, this limitation would not be present, and a higher number of particles up to $N=20$ can be expected to be accessible when no additional symmetries are used. If, for instance, the total magnetization is still a valid quantum number in the presence of the perturbation, and when the observable we are interested in does not couple different sectors, our method should also be able to go beyond $N=20$.

Finally, it should be stressed once more that the method we propose is applicable to a larger scope of problems than was shown here for the example of the central spin model. 
Particular examples one could think of are perturbations of the integrable XXX or XXZ spin chain with a transverse field, or driving one of the spins in the chain with a time-dependent field. Another possibility is to make a local change in the anisoptropy of the XXZ chain, and study integrability breaking effects on the conserved charges.
Using the Bethe states as the computational basis then facilitates an efficient computation of the time evolution of these quantities, since their corresponding operators are diagonal in this particular basis, and have matrix elements which are simple functions of the rapidities of the Bethe states.
We will consider the application of our method to the XXZ spin chain in future publications. In addition we hope to apply this method to study perturbations of the Lieb-Liniger model with an eye to applications to quenches in one dimensional Bose
gases.  We think this approach will act as a useful complement to the numerical renormalization group methodology used to study such quenches in Refs. 
\cite{2012_Caux_PRL_109,2014_Brandino_arXiv}.  

We thank B. Wouters, A. Faribault, D. Schuricht, R. Hanson and L. Vandersypen for useful discussions, and N. Renaud for illustrations. This work was supported by the Netherlands Organisation for Scientific Research (NWO) and the Foundation for Fundamental Research on Matter (FOM), and forms part of the activities of the Delta-Institute for Theoretical Physics (D-ITP). This research was done in part under the auspices of the CMPMS Dept. at Brookhaven National Laboratory, which in turn is supported by the U.S. Department of Energy, Office of Basic Energy Sciences, under Contract No. DE-AC02-98CH10886. O.E.A. is supported by the Swiss National Science Foundation.

\bibliography{literature}

\end{document}